\documentclass[sn-basic, authoryear]{sn-jnl}

\usepackage{amsmath}
\usepackage{multicol}
\usepackage{amsfonts}
\usepackage{textcomp}
\usepackage{stfloats}
\usepackage{url}
\usepackage{verbatim}
\usepackage{graphicx}

\usepackage{indentfirst}
\usepackage{psfrag,epsf}

\usepackage{enumerate}

\RequirePackage{amsthm}
\usepackage{amssymb}
\usepackage{booktabs}
\usepackage{bbm}
\usepackage{color}
\usepackage{bm}
\usepackage{siunitx}
\usepackage{subfigure}

\usepackage{makecell}
\usepackage{arydshln}
\usepackage{diagbox}
\usepackage{multirow}
\usepackage{array}
\usepackage{float}
\usepackage{cleveref}

\usepackage{algorithm}
\usepackage{algorithmic}
\usepackage{tikz}

\usepackage{threeparttable}
\usetikzlibrary{decorations.pathreplacing,calligraphy}
\usetikzlibrary{arrows,positioning,fit,shapes,calc}

\usepackage{dcolumn}%

\newtheorem{thm}{Theorem}[section]
\newtheorem{Example}{Example}

\newcommand{\bee} {\begin{equation} }
\newcommand{\ene}{\end{equation}}

 \numberwithin{equation}{section}
 \numberwithin{lem}{section}

\DeclareGraphicsExtensions{.pdf,.png,.jpg}
\def\qed{{\hfill $\Box$\medskip}}

\makeatletter%
\@addtoreset{equation}{section}

\newcommand\figcaption{\def\@captype{figure}\caption}
\newcommand\tabcaption{\def\@captype{table}\caption}
\makeatother%

\begin{document}




\title[Article Title]{Transfer Learning in Regression with Influential Points}


\author[1]{\fnm{Bingbing} \sur{Wang}} 
\email{bbwangstat1@stu.suda.edu.cn}
\author[1]{\fnm{Jiaqi} \sur{Wang}} 
\email{20234007011@stu.suda.edu.cn}
\author*[2]{\fnm{Yu} \sur{Tang}} 
\email{ytang@suda.edu.cn}

\affil[1]{\orgdiv{School of Mathematical Sciences}, \orgname{Soochow University}, \orgaddress{\city{Suzhou}, \postcode{215031}, \state{Jiangsu}, \country{China}}}

\affil[2]{\orgdiv{School of Future Science and Engineering}, \orgname{Soochow University}, \orgaddress{\city{Suzhou}, \postcode{215006}, \state{Jiangsu}, \country{China}}}

\abstract{
Regression prediction plays a crucial role in practical applications and strongly relies on data annotation. However, due to prohibitive annotation costs or domain-specific constraints, labeled data in the target domain is often scarce, making transfer learning a critical solution by leveraging knowledge from resource-rich source domains. In the practical target scenario, although transfer learning has been widely applied, influential points can significantly distort parameter estimation for the target domain model. This issue is further compounded when influential points are also present in source domains, leading to aggravated performance degradation and posing critical robustness challenges for existing transfer learning frameworks. In this study, we innovatively introduce a transfer learning collaborative optimization (Trans-CO) framework for influential point detection and regression model fitting. Extensive simulation experiments demonstrate that the proposed Trans-CO algorithm outperforms competing methods in terms of model fitting performance and influential point identification accuracy. Furthermore, it achieves superior predictive accuracy on real-world datasets, providing a novel solution for transfer learning in regression with influential points.}


\maketitle

\section{Introduction}\label{section:1}
Regression prediction is of great importance in many practical application situations, and the training of regression models is highly dependent on data annotation. Inaccurate or incomplete data annotation can lead the model to assimilate erroneous information, thereby affecting the accuracy of predictions. However, in numerous practical applications, due to the high cost of obtaining labeled data in the target domain or restrictions in specific fields, the amount of labeled data in the target domain is often limited, making it difficult to effectively train the target model. Transfer learning addresses this challenge by integrating a large amount of data from related domains, alleviating the problem of data scarcity in the target domain. 

Transfer learning has been drawing increasing focus in many fields recently, and numerous scholars have undertaken extensive work. Li et al. (\citeyear{li2022transfer}) introduced a data-driven transfer learning method termed Trans-Lasso for high-dimensional linear regressions, which involves meticulously constructing the candidate estimators and selecting an auxiliary set via the $l_q$-distance under vanishing-difference assumption. Tian and Feng (\citeyear{tian2023transfer}) focused on transfer learning for high-dimensional Generalized Linear Models (GLMs). Li et al. (\citeyear{li2024estimation}) also put forward a transfer learning algorithm called TransHDGLM for high-dimensional GLMs. Jin et al. (\citeyear{jin2024transfer}) developed the Trans-Lasso QR method specifically designed for high-dimensional quantile regression. Chen and Song (\citeyear{chen2025transfer}) introduced a new transfer learning approach for the semiparametric varying coefficient spatial autoregressive models, enabling efficient knowledge transfer from source data to the target model. Lou and Yang (\citeyear{lou2025joint}) employed transfer learning techniques in combination with historical data to estimate the predicted values for the current period. Tripuraneni et al. (\citeyear{tripuraneni2021provable}) assumed in multi-task linear regression that source models and target model share a common low-dimensional linear representation for transfer learning. Lin et al. (\citeyear{lin2024profiled}) proposed the Profiled Transfer Learning (PTL) estimator for transfer learning under the flexible approximate-linear assumption, enabling an arbitrarily large difference between target and source parameters measured by $\|\bm{\beta}-\bm{\beta}_{(k)}\|_{q}$. 
Although transfer learning has been widely applied in regression tasks, the presence of influential points in both source and target data can severely undermine the effectiveness of these transfer learning methods.

Influential points refer to data points that exert a significant impact on parameter estimation, prediction results, or the goodness-of-fit of a statistical model. An influential point does not necessarily exhibit numerical outliers; instead, it influences the model through its relationships with other variables within the datasets (Aguinis et al. \citeyear{aguinis2013best}), making it difficult to be directly removed in advance. The impact of such data points is often imperceptible through direct examination of raw data, yet such points warrant further investigation either because they may be in error or because of their differences from the rest of the data (Belsley et al. \citeyear{belsley2005regression}). It should be noted that the identification of influential points is essentially an open statistical challenge, and how to determine influential points is actually a difficult task to clarify. 
For the detection of an individual influential point, classic leave-one-out methods such as Cook's distance and DFFITS can be employed (Cousineau and Chartier \citeyear{cousineau2010outliers}). However, when multiple influential points coexist, their mutual interference may lead to masking or swamping, significantly increasing the difficulty of detection. In such cases, it is necessary to incorporate more robust statistical methods. Klivans et al. (\citeyear{klivans2018efficient}) proposed a polynomial-time algorithm to perform linear or polynomial regression that is resilient to adversarial corruptions. She and Owen (\citeyear{she2011outlier}) introduced a thresholding based iterative procedure for outlier detection, and they found that the hard threshold version, which satisfies some nonconvex criteria, can properly identify multiple outliers in some challenging cases.
Liu et al. (\citeyear{liu2020high}) proposed a filtering algorithm that incorporates a novel stochastic outlier removal technique for robust sparse mean estimation. Bottmer et al. (\citeyear{bottmer2022sparse}) obtained a sparse and cellwise robust regression method that is resistant to outliers in the cells of the data matrix by employing sparse shooting with a simple sparse robust estimator. However, these methods may not perform satisfactorily when the data is insufficient. 
Yan et al. (\citeyear{yan2024comprehensive}) conducted a systematic summary of the deep transfer learning methods and frameworks employed in the field of industrial time series anomaly detection, encompassing models such as Convolutional Neural Networks (CNN) (Yao et al. \citeyear{yao2022model}; Pan et al. \citeyear{pan2023transfer}), Fully Convolutional Networks (FCN) (Lockner et al. \citeyear{lockner2022transfer}), and Long Short-Term Memory networks (LSTM) (Zabin et al. \citeyear{zabin2023hybrid}; Abdallah et al. \citeyear{abdallah2023anomaly}; Panjapornpon et al. \citeyear{panjapornpon2023explainable}). However, these deep learning methods often entail an extremely large computational load during the training phase, and the model parameters are usually in a latent state, making them difficult to interpret intuitively. In contrast, in those domains with which we are more familiar, traditional regression models have already demonstrated favorable application performance.  It is worth mentioning that, as of now, research on transfer learning for regression problems containing influential points remains extremely scarce.

In this paper, we study transfer learning within the context of regression amidst the presence of influential points, and introduce a corresponding transfer learning algorithm. The novel contributions of this study are summarized as follows:

\begin{itemize}
    \item To improve the performance of regression models in scenarios where data contain influential points, especially when tackling the challenge of limitations imposed by insufficient data volume, transfer learning techniques are introduced in this paper. To address transfer learning for influential point detection in regression models, we propose an algorithm named Trans-CO, which utilizes parameter knowledge from the source model to transfer parameters to the target model.
    \item Our Trans-CO method achieves robust performance not only in conventional statistical modeling where $n<p$, but also extends seamlessly to high-dimensional regimes where $n>p$ by leveraging the same penalty structure that adaptively balances sparsity and estimation accuracy.
    \item Comparative experiments are conducted to assess three methods under different sample sizes, variable sparsity, and to examine how drift proportions and source model count affect transfer learning. Simulations are also run under heteroscedasticity and when the unique identification conditions are not met. Both simulation and real data analysis indicated that our proposed Trans-CO method outperformed the others in multiple aspects.
\end{itemize}

The remainder of this paper is structured as follows. In Sect. \ref{section:Methodology}, We initially introduced a robust regression model tailored for influential point detection, along with a methodology to select optimal parameters using the Bayesian Information Criterion (BIC). Subsequently, under the assumption of linear approximation, we proposed an algorithm named Trans-CO specifically designed for transfer learning in the context of influential point detection within regression models. Simulation experiments are reported in Sect. \ref{Simulation experiments}, and real data analysis is presented in Sect. \ref{Experiments on Real Data}. In Sect. \ref{section:Conclusion and Discussion}, we review our contributions and outline promising directions for future work.


\section{Methodology}
\label{section:Methodology}
\subsection{Robust regression model for influential point detection}

Consider the following mean-shift model that allows any observation as an influential point:
\begin{equation}
\label{equation:1}
    Y_i = \bm{X}_i\bm{\beta}+\gamma_i+\epsilon_i,\quad i=1,...,n,
\end{equation}
where $Y_i$ represents the response variable, $\bm{X}_i\in \mathbb{R}^p$ is the vector of regression covariates, $\bm{\beta}\in \mathbb{R}^p$ denotes the coefficient vector, $\gamma_i$ is non-zero if observation $i$-th is an influential point, and $\epsilon_i$ is the random error satisfying $E(\epsilon_i)=0$ and $E(\epsilon_i^2)=\sigma^2$. For all $n$ observations, we integrate the above model as
\begin{equation}
    \bm{Y} = \bm{X\beta+\gamma+\epsilon}, \quad \bm{\epsilon} \sim \mathcal{N}(0,\sigma^2\bm{I}),
\end{equation}
where $\bm{Y}=(Y_1,...,Y_n)^{\top}\in \mathbb{R}^{n}$, $\bm{X}=(\bm{X}_1,...,\bm{X}_n)^{\top}\in \mathbb{R}^{n\times p}$, $\bm{\gamma}=(\gamma_1,...,\gamma_n)^{\top}\in\mathbb{R}^{n}$, and $\bm{\epsilon}\in\mathbb{R}^{n}$ is the random error vector.
It contains $p+n$ regression parameters, encompassing $\bm{\beta}$ and $\bm{\gamma}$. In the presence of multiple influential points, the estimates of the ordinary least squares (OLS) parameter for linear models are significantly biased. The mean-shift model is constructed to visually evaluate the magnitudes of influential points while simultaneously obtaining robust regression coefficients. It is fitted by enforcing sparsity on $\bm{\gamma}$, with the aim of attaining a more precise estimation of $\bm{\beta}$ where multiple influential points exist.

She and Owen (\citeyear{she2011outlier}) proposed an algorithm called a thresholding (denoted $\Theta$) based iterative procedure for outlier detection ($\Theta$-IPOD). Parameter estimation is conducted by minimizing the following objective function:
 \begin{equation}
 \label{opt3}
     f_P(\bm{\beta},\bm{\gamma})\equiv \frac{1}{2}\|\bm{Y}-\bm{X}\bm{\beta}-\bm{\gamma} \|_2^2  + \sum\limits_{i=1}^{n}P(\gamma_i;\lambda_i),
 \end{equation}
 where $\lambda_i$ are a collection of penalty parameters. A threshold function $\Theta(\cdot; \lambda)$ is coupled with nonconvex penalty $P(\gamma;\lambda) =P(0;\lambda)+P_{\Theta}(\gamma;\lambda)+q(\gamma;\lambda)$, where $ P_{\Theta}(\gamma;\lambda) = \int_0^{\left|\gamma\right|}(sup\{t:\Theta(t;\lambda)\leq u\}-u) du$, $q(\cdot;\lambda)$ is nonnegative and $q(\Theta(\gamma;\lambda);\lambda)=0$ for all $\gamma$. $\Theta(\gamma; \lambda)$ is an odd monotone unbounded shrinkage rule for $\gamma$, at any $\lambda$.

\begin{algorithm}[ht]
    \caption{Robust regression learner $\Theta$-IPOD}
    \label{alg: IPOD}
    \renewcommand{\algorithmicrequire}{\textbf{Input:}}
    \renewcommand{\algorithmicensure}{\textbf{Output:}}
    \begin{algorithmic}[1]
        \scriptsize{\REQUIRE $\bm{X} \in \mathbb{R}^{n\times p}$; $\bm{Y} \in \mathbb{R}^{n}$; penalty parameters $\bm{\lambda}$; relative iterative converence tolerance $\epsilon$; a threshold function $\Theta(\cdot;\cdot)$ 
        \ENSURE A robust estimate $\hat{\bm{\beta}},\hat{\bm{\gamma}}$  
        \STATE Initialize $\bm{\gamma}^{(0)}$, $i=0$ , $converged \leftarrow False$
        \STATE $\bm{H} = \bm{X}(\bm{X}^T\bm{X})^{-1}\bm{X}^T, \bm{r} = \bm{Y} - \bm{HY}$
        \WHILE {not $converged$}
            \STATE $\bm{\gamma}^{(i+1)} \leftarrow \Theta(\bm{H}\bm{\gamma}^{(i)} + \bm{r}; \bm{\lambda})$
            \IF{$\| \bm{\gamma}^{(i+1)} - \bm{\gamma}^{(i)} \|_\infty < \epsilon$}
                \STATE $converged \leftarrow True$
            \ENDIF
            \STATE $i \leftarrow i + 1$
        \ENDWHILE
        \STATE $\hat{\bm{\gamma}} = \bm{\gamma}^{(i)}, \hat{\bm{\beta}} = (\bm{X}^T\bm{X})^{-1}\bm{X}^T(\bm{Y} - \hat{\bm{\gamma}})$
        \RETURN $\hat{\bm{\beta}},\hat{\bm{\gamma}}$}
    \end{algorithmic}
\end{algorithm}
The algorithm \ref{alg: IPOD} employs an alternating optimization approach. 
Fixed $\bm{\gamma}$, $\bm{\beta}$ is the OLS estimate of $\bm{Y}-\bm{\gamma}$ regressed on $\bm{X}$. 
Fixed $\bm{\beta}$, $\bm{\gamma}$ is obtained using the threshold $\Theta(\bm{y-X\beta};\bm{\lambda})$ to ensure the convergence of the objective function in iterations, where $\bm{\lambda} = (\lambda_1,...,\lambda_n)$. For simply, $\bm{\gamma}$ can be updated via:
\begin{equation}
    \bm{\gamma}^{(j+1)}=\Theta(\bm{Y}-\bm{X}(\bm{X}^T\bm{X})^{-1}\bm{X}^T(\bm{Y}- \bm{\gamma}^{(j)});\bm{\lambda})=\Theta(\bm{Y}-\bm{H}\bm{Y}+ \bm{H}\bm{\gamma}^{(j)};\bm{\lambda}).
\end{equation}
This algorithm \ref{alg: IPOD} necessitates a preliminary regression step; however, it outperforms the preliminary regression approach. 

For example, the following hard-thresholding rule satisfies the definition of the thresholding function $\Theta$: 
\begin{equation}
    \Theta_{hard}(\gamma;\lambda)=\begin{cases}
    0, & |\gamma|\leq\lambda,  \\
    \gamma, & |\gamma|>\lambda. 
\end{cases}
\end{equation} When $\bm{\gamma}$ and $\bm{\lambda}$ are high-dimensional vectors, element-wise hard-thresholding is applied to corresponding parameter positions. Moreover, it identifies that certain nonconvex criteria are capable of accurately detecting multiple influential points, thereby effectively mitigating the masking (where actual influential points are overlooked) and swamping (where noninfluential points are falsely identified) phenomena in influential point detection.

Subsequently, regarding the selection of $\lambda_i$, set $\lambda_i=\lambda_{adj}\sqrt{1-h_i}$, where $h_i$ is the $i$-th diagonal element of $\bm{H}$, and the regularization parameter $\lambda_{adj}$ is tuned using the Bayesian Information Criterion (BIC) for parameter selection.
$\lambda_{adj}$ is adjusted over a range that spans from $\|(\bm{I}-\bm{H})\bm{Y}.\sqrt{diag(\bm{I}-\bm{H})}\|_{\infty}
$ to $0$.
Here, $nz(\lambda_{adj})$ is defined as the set of indices $\{i:\hat{\bm{\gamma}}(\lambda_{adj})\neq 0\}$, corresponding to the non-zero components of the estimated vector $\hat{\bm{\gamma}}(\lambda_{adj})$, and the degrees of freedom are given by $\text{DF}(\lambda_{adj})=|nz(\lambda_{adj})|$. A slightly modified version of the BIC is then employed as follows:
\begin{equation}\label{bic1}
    \text{BIC}^*(\lambda_{adj})=m\mathrm{log}(\text{RSS}/m)+q(\mathrm{log}(m)+1),
\end{equation}
where $m=n-p$, $\text{RSS}=\|(\bm{I}-\bm{H})(\bm{Y}-\hat{\bm{\gamma}})\|_2^2$ and $q=\text{DF}(\lambda_{adj})+1$.

However, in high-dimensional spaces, the parameters obtained by the aforementioned algorithms may not be particularly accurate. By leveraging transfer learning, the model can more efficiently utilize limited data to achieve more accurate parameter estimation.
\subsection{Transfer learning
collaborative optimization framework}
We consider mean-shift models for both target and $K$ sources in this paper. The target dataset $\{(\bm{X}_i,Y_i)\}_{i=1}^{n}$ and the source datasets for each domain $k=1,...,K$, denoted as $ \{(\bm{X}_{j(k)},Y_{j(k)})\}_{j=1}^{N_{(k)}}$, are individually assumed to be independently and identically distributed (i.i.d.). The target dataset is generated from model (\ref{equation:1}). Meanwhile, for $k=1,...,K$, the source data is generated by the following model:
\begin{equation}
\label{equation:2}
    \bm{Y}_{(k)} = \bm{X}_{(k)}\bm{\beta}_{(k)}+\bm{\gamma}_{(k)}+\bm{\epsilon}_{(k)}, \quad \bm{\epsilon}_{(k)} \sim \mathcal{N}(0,\sigma_{(k)}^2\bm{I}),
\end{equation}
where $\bm{Y}_{(k)} = (Y_{1(k)},...,Y_{N_{(k)}(k)})^{\top}\in \mathbb{R}^{N_{(k)}}$, $\bm{X}_{(k)}=(\bm{X}_{1(k)},...,\bm{X}_{N_{(k)}(k)})^{\top}\in \mathbb{R}^{N_{(k)}\times p}$, $\bm{\gamma}_{(k)}=(\gamma_{1(k)},...,\gamma_{N_{(k)}(k)})^{\top}\in\mathbb{R}^{N_{(k)}}$, $\bm{\beta}_{(k)}\in\mathbb{R}^{p}$ represents the regression coefficient of each source model and $\bm{\epsilon}_{(k)}\in\mathbb{R}^{N_{(k)}}$ denotes the random error.

To incorporate information from the source data, we adopt the following approximate-linear assumption imposed by Lin et al. (\citeyear{lin2024profiled}): 
\begin{equation}
\label{assumption:approximate-linear}
    \bm{\beta} = \bm{B}\bm{w}+\bm{\delta}.
\end{equation}
Here, $\bm{w}=(w_1,...,w_K)^{\top}\in\mathbb{R}^{K}$ denotes the weight vector assigned to coefficients of $\bm{B}=(\bm{\beta}_{(1)},...,\bm{\beta}_{(K)})\in\mathbb{R}^{p\times K}$ in source models. We assume that the regression coefficients $\bm{\beta}_{(k)}$ are linearly independent across different $k$ in this paper. The residual vector $\bm{\delta}\in\mathbb{R}^{p}$ is ideally small and sparse. However, this approximate-linear assumption allows the difference between the target and source coefficients (i.e. $\|\bm{\beta}-\bm{\beta}_{(k)}\|_{q}$) to be arbitrarily large. Given this assumption, transferring the regression coefficients $\bm{\beta}_{(k)}$ to the target domain requires only the estimation of two components: the weight vector $\bm{w}$ and the residual term $\bm{\delta}$.

To address this challenge, Lin et al. (\citeyear{lin2024profiled}) proposed that under assumption $\bm{\beta}_{(k)}\Sigma\bm{\delta}=0$ for each $k=1,...,K$, both $\bm{w}$ and $\bm{\delta}$ are uniquely identifiable. Moreover, they put forward the Profiled Transfer Learning (PTL) estimator, a two-step procedure: $\bm{w}$ is first estimated via regression of $\bm{Y}$ on $\bm{X}\hat{\bm{B}}$, where $\hat{\bm{B}}$ is the estimator of $\bm{B}$ through $\Theta$-IPOD. $\bm{\delta}$ is subsequently estimated using LASSO regression on a profiled response $\bm{e} =\bm{Y}-\bm{X}\hat{\bm{B}}\hat{\bm{w}}$ derived from the first-step residuals. 
 
Inspired by this work, we propose a transfer learning framework under the mean-shift model that leverages information from source models to optimize two objectives for the target model: enhancing influential point detection performance and improving the robustness of regression coefficient estimation. Thus, we consider the target model of transfer learning collaborative optimization:
\begin{equation}
    \bm{Y} = \bm{XBw+X\delta+\gamma+\epsilon}, \quad \bm{\epsilon} \sim \mathcal{N}(0,\sigma^2\bm{I}).
\end{equation}

Our framework aims to minimize the following objective function:
\begin{equation}
\label{objective function}
    f(\bm{\gamma},\bm{w},\bm{\delta})=\frac{1}{2}\|\bm{Y}-\bm{X}\hat{\bm{B}}\bm{w-\bm{X}\delta-\gamma}\|_2^2+\sum\limits_{j=1}^{p}P(\delta_j;\lambda) +\sum\limits_{i=1}^{n}P(\gamma_i;\lambda).
\end{equation}
Notably, our transfer learning framework involves three key parameters requiring estimation: the weight vector $\bm{w}$, the residual term $\bm{\delta}$ and the influential point detection parameter $\bm{\gamma}$. We employ an alternating iterative optimization procedure to estimate these parameters, with each iteration comprising two sequential steps, and the detailed optimization workflow is summarized as follows:

1) Ordinary Least Squares (OLS) estimation of the weight vector $\bm{w}$:
\begin{equation}
    \bm{w}^{(i+1)}=(\bm{Z}^{\top}\bm{Z})^{-1}\bm{Z}^{\top}(\bm{Y}-\bm{X}\bm{\delta}^{(i)}-\bm{\gamma}^{(i)}),
\end{equation}
where $\bm{Z} = \bm{X}\hat{\bm{B}}$. This leads to residuals $\bm{e} = \bm{Y -X\hat{B}}\bm{w}^{(i+1)}$. 
Note that $\bm{\delta}$ and $\bm{\gamma}$ are assumed to be sparse, we can
directly work in an augmented data space $\bm{e}^{(i+1)} = \bm{M\xi}$, where $\bm{M} = \begin{bmatrix} 
\bm{X} & \bm{I}_{n\times n} 
\end{bmatrix}$ and $\bm{\xi} = \begin{bmatrix} 
\bm{\delta} & \bm{\gamma} 
\end{bmatrix}^{\top}$. 
Similarly to the thresholding-based iterative selection procedures (TISP) constructed for non-orthogonal regression matrices proposed by She (\citeyear{she2009thresholding}), we next employ an analogous rationale to perform selection on $\bm{\xi}$.

2)Estimation of $\bm{\xi} = \begin{bmatrix} 
\bm{\delta} & \bm{\gamma} 
\end{bmatrix}^{\top}$ through satisfaction of the hard-thresholding function condition:
\begin{equation}
    \bm{\delta}^{(i+1)}=\Theta_{hard}(\bm{\delta}^{(i)}+\frac{\bm{X}^{\top}}{k_0^2}(\bm{Y-X\hat{B}}\bm{w}^{(i+1)}-\bm{X}\bm{\delta}^{(i)}-\bm{\gamma}^{(i)});\frac{\lambda}{k_0^2}),
\end{equation}

\begin{equation}
    \bm{\gamma}^{(i+1)}=\Theta_{hard}(\bm{\gamma}^{(i)}+\frac{1}{k_0^2}(\bm{Y-X\hat{B}}\bm{w}^{(i+1)}-\bm{X}\bm{\delta}^{(i)}-\bm{\gamma}^{(i)});\frac{\lambda}{k_0^2}),
\end{equation}
where $k_0=\sigma_{max}(\bm{M})+1$, representing the max singular value of $\bm{M}$. The advantage of Our Trans-CO lies in leveraging the Oracle property of nonconvex penalties, which enables accurate identification of truly nonzero $\bm{\xi}$ while achieving the convergence for parameter estimates. In addition, the convergence property of our Trans-CO holds not only under the general condition of $n>p$, but also extends to the high-dimensional scenario where $n<p$.

Suppose that the spectral decomposition of the hat matrix $\bm{H_z} = \bm{Z}(\bm{Z}^{\top}\bm{Z})^{-1}\bm{Z}^{\top}$ is given by $\bm{H_z} =\bm{UDU}^{\top}$, and let $\bm{U}_c\in\mathbb{R}^{n\times (n-K)}$ consist of the columns of $\bm{U}$ indexed by $c = \{i:D_{ii}=0\}$. Define $\bm{C} = \bm{U}_c^{\top}\bm{M}\bm{M}^{\top}\bm{U}_c$. Given that $\bm{M}\bm{M}^{\top} = \bm{X}\bm{X}^{\top}+\bm{I}$ is positive definite and the columns of $\bm{U}_c$ are linearly independent, $\bm{C}$ is symmetric positive definite and has a unique inverse square root $\bm{C}^{-\frac{1}{2}}$.
Since $\bm{w}$ is estimated through OLS,
we can obtain a $\bm{Zw}$-eliminated version of the target model:
\begin{equation}\label{modelall}
    \tilde{\bm{Y}} = \bm{A\xi}+\bm{\epsilon}',\quad \bm{\epsilon}' \sim \mathcal{N}(0,\sigma^2\bm{C}^{-1}).
\end{equation}
where $\tilde{\bm{Y}} = \bm{P}\bm{Y}$, $\bm{A}=\bm{P}\bm{M}$, $\bm{\epsilon}' = \bm{P}\bm{\epsilon}$, and $\bm{P} = \bm{C}^{-\frac{1}{2}}\bm{U}_c^{\top}$. Consequently, it readily follows that $\bm{A}\bm{A}^{\top}=\bm{I}$. This demonstrates that any statistical model constructed with a non-orthogonal design matrix can be equivalently transformed into a model based on an orthogonal matrix through a linear transformation. This simplified form is also beneficial for finding optimal parameter $\lambda$ by constructing BIC criteria.

Given a penalty parameter $\lambda$ tuned over a sufficiently broad range, we compute the corresponding estimate $\hat{\bm{\xi}}(\lambda)$ and define the model degrees of freedom as $\text{DF}_{\xi}(\lambda) = |\{i:\hat{\bm{\xi}}(\lambda)\neq0\}|$ where the cardinality operator counts the number of non-zero coefficients. In order to effectively balance the goodness of fit and model complexity, we give the correct form of BIC similar to (\ref{bic1}) relying on model (\ref{modelall}):
\begin{equation}
    \text{BIC}^*(\lambda)=m\mathrm{log}(\text{RSS}/m)+q(\mathrm{log}(m)+1),
\end{equation}
where $m=n-K$, $\text{RSS}=\|\tilde{\bm{Y}}-\bm{A}\hat{\bm{\xi}}\|_2^2$ and $q=\text{DF}_{\xi}(\lambda)+1$. The optimal penalty parameter $\lambda$ is determined by minimizing the BIC$^*$ criterion.


\begin{thm}
\label{thm:1}
$\Theta(\xi; \lambda)$ is an odd monotone unbounded shrinkage rule for $\xi$, at any $\lambda$, and let  corresponding penalty $P(\xi;\lambda)$ follows the definition in the equation (\ref{opt3}). The condition $\bm{B}^{\top}\bm{\Sigma}\bm{\delta}=0$ is guaranteed for unique identification. The objective function is defined by equation (\ref{objective function}). Then the Trans–OD iteration sequence $(\bm{\xi}^{(i)},\bm{w}^{(i)})$ satisfies
\begin{equation}\label{equation objective function}
\begin{aligned}
    f(\bm{\xi}^{(i)},\bm{w}^{(i)})\geq f(\bm{\xi}^{(i+1)},\bm{w}^{(i)}) \geq f(\bm{\xi}^{(i+1)},\bm{w}^{(i+1)}).
\end{aligned}
\end{equation}
\end{thm}
The proof of this theorem is shown in the Appendix \ref{Proof1}. This conclusion provides critical theoretical guarantees for the iterative optimization of model parameters in transfer learning scenarios, demonstrating that the proposed algorithm converges to a stable solution within the objective function space. Eventually the estimate of $\bm{\beta}$ can be obtained by $\hat{\bm{\beta}} = \hat{\bm{B}}\hat{\bm{w}}+\hat{\bm{\delta}}$. Algorithm \ref{alg: IPODTR} summarizes the proposed transfer learning framework. The empirical performance of the algorithm is subsequently validated through both simulation studies and real-world case analyses in the following sections.


\begin{algorithm}[ht]
    \caption{Transfer Learning Collaborative Optimization (Trans-CO)}
    \label{alg: IPODTR}
    \renewcommand{\algorithmicrequire}{\textbf{Input:}}
    \renewcommand{\algorithmicensure}{\textbf{Output:}}
    \begin{algorithmic}[1]
        {\REQUIRE $\bm{X}_{(k)} \in \mathbb{R}^{N_{(k)}\times p}$; $\bm{Y}_{(k)} \in \mathbb{R}^{N_{(k)}}$; $k=1,...,K$; $\bm{X} \in \mathbb{R}^{n\times p}$; $\bm{Y} \in \mathbb{R}^{n}$; penalty parameters $\bm{\lambda}$; relative iterative converence tolerance $\epsilon$; a hard-threshold function $\Theta_{hard}(\cdot;\cdot)$ 
        \ENSURE A robust transfer leaning estimator $\hat{\bm{\beta}}$  
        \FOR{$k=1$ to $K$}
            \STATE $\hat{\bm{\gamma}}_{(k)} \leftarrow IPOD(\bm{X}_{(k)},\bm{Y}_{(k)})$
            \STATE$\hat{\bm{\beta}}_{(k)} \leftarrow OLS(\bm{X}_{(k)},\bm{Y}_{(k)}-\hat{\bm{\gamma}}_{(k)})$ 
        \ENDFOR
        \STATE $\bm{Z} \leftarrow \bm{X}\hat{\bm{B}}$
        \STATE Initialize $i\leftarrow0$, $\hat{\bm{\beta}}^{(i)} \leftarrow OLS(\bm{X,Y})$, $\hat{\bm{\gamma}}^{(i)} \leftarrow \bm{Y} - \bm{X}\bm{\beta}^{(i)}$, $\bm{\delta}^{(i)}\leftarrow\bm{0}$, $converged \leftarrow False$ \# LassoCV can be used instead of OLS above in high-dimensional regression when $n<p$.
        \WHILE {not $converged$}
            \STATE $ \hat{\bm{w}}^{(i+1)} \leftarrow (\bm{Z}'\bm{Z})^{-1}\bm{Z}'(\bm{Y} - \bm{\gamma}^{(i)}-\bm{X}\bm{\delta}^{(i)})$
            \STATE $\bm{\delta}^{(i+1)}\leftarrow\Theta_{hard}(\bm{\delta}^{(i)}+\frac{\bm{X}^{\top}}{k_0^2}(\bm{Y-X\hat{B}}\bm{w}^{(i+1)}-\bm{X}\bm{\delta}^{(i)}-\bm{\gamma}^{(i)});\frac{\lambda}{k_0^2})$
            \STATE $\bm{\gamma}^{(i+1)}\leftarrow\Theta_{hard}(\bm{\gamma}^{(i)}+\frac{1}{k_0^2}(\bm{Y-X\hat{B}}\bm{w}^{(i+1)}-\bm{X}\bm{\delta}^{(i)}-\bm{\gamma}^{(i)});\frac{\lambda}{k_0^2})$
            \STATE $\hat{\bm{\beta}}^{(i+1)} \leftarrow \bm{B}\hat{\bm{w}}^{(i+1)}+\hat{\bm{\delta}}^{(i+1)}$
            \IF{$\| \bm{\gamma}^{(i+1)} - \bm{\gamma}^{(i)} \|_\infty < \epsilon$}
                \STATE $converged \leftarrow True$
            \ENDIF
            \STATE $i \leftarrow i + 1$
        \ENDWHILE
        \STATE $\hat{\bm{\gamma}} \leftarrow \bm{\gamma}^{(i+1)}, \hat{\bm{w}} \leftarrow \bm{w}^{(i+1)}, \hat{\bm{\delta}} \leftarrow \bm{\delta}^{(i+1)}, \hat{\bm{\beta}} \leftarrow \hat{\bm{B}}\hat{\bm{w}}+\hat{\bm{\delta}}$ 
        \RETURN $\hat{\bm{\beta}}$}
    \end{algorithmic} 
\end{algorithm}


\section{Simulation experiments}\label{Simulation experiments}
To evaluate the practical efficacy of our proposed transfer learning algorithm, we perform simulation experiments that objectively measure its performance across diverse operational scenarios. 
Specifically, we examine the performance of
three estimators: (1) the profiled transfer leaning (PTL) estimator, (2) the $\Theta$-IPOD estimator using the target data only, and (3) our Trans-CO estimator. 

To quantify the accuracy of parameter estimation, we adopt the following metric as an error measure:
\begin{equation}
    \text{MSE}=\frac{1}{p}\|\hat{\bm{\beta}}-\bm{\beta}\|_2^2.
\end{equation}
In addition, we use F1-score to evaluate the detection accuracy of influential points:
\begin{equation}
    \text{F1-score}=\frac{2P_{ppv} \cdot P_{tpr}}{P_{ppv} + P_{tpr}},
\end{equation}
where the positive predicted value $P_{ppv} = \frac{TP}{TP+FP}$ also known as precision refers to the proportion of correctly detected true influential points among the total samples detected as influential points, and the true positive rate $P_{tpr} = \frac{TP}{TP+FN}$ also known as recall is the proportion of correctly detected true influential points among the true influential points. F1-score combines precision and recall, balancing the performance of the model through harmonic averaging.
\begin{Example}
\label{ex1}
This is an example revised from Li et al. (\citeyear{li2022transfer}), Tripuraneni et al. (\citeyear{tripuraneni2021provable}) and Lin et al. (\citeyear{lin2024profiled}). 
\begin{itemize}
\item The target data $\{(\bm{X}_i,Y_i)\}_{i=1}^{n}$ and source data $ \{(\bm{X}_{j(k)},Y_{j(k)})\}_{j=1}^{N_{(k)}}$ are i.i.d. observations generated from the model (\ref{equation:1}) and the model (\ref{equation:2}) respectively, where $\bm{X}_i\sim N(0,\Sigma)$, $\bm{X}_{j(k)}\sim N(0,\Sigma_{(k)})$, $\Sigma_{(k)} = \Sigma = \bm{I}_p$, and $\sigma^2_k = \sigma^2 = 1$ for $ k=1,...,K$. The sample size of the target data $n\in\{150, 200, 300\}$. Set the number of source datasets $K=5$, the weight vector $\bm{w}=(3/2, 3/4, 0, 0, -5/4)^{\top}$, and for $k=1,...,5$, the sample size of $K$ source datasets to be the same, denoted as $N_{(k)}=N \in \{1000, 1500, 2000\}$. We consider $p=100$, $h=6$, $\rho=0.1$.
\item Generate $\bm{B}$. Set $r_0 = \lfloor s/3 \rfloor$ and $s_{\delta} = \lfloor s/5 \rfloor$ with $s \in\{ 25 (sparse) ,75 (dense)\}$. Let $\Omega \in \mathbb{R}^{r_0 \times K}$ be a matrix with elements generated from $N(0,1)$, and let $U_K = (u_1, \ldots, u_K)\in \mathbb{R}^{r_0 \times K}$ be the first $K$ left singular vectors of $\Omega$ obtained from its singular value decomposition (SVD), then $\bm{B} = (2U_K,0.3\bm{I}_{s-r
_0,K},\bm{0}_{p-s,K})\in \mathbb{R}^{p \times K}$.
\item Generate $\bm{\delta}$. Let $S$ be an index set with $|S| = s_{\delta}$ randomly sampled from $\{s + 1, . . . , p\}$ without replacement. Next, for each $j \in S$, generate $\delta_j$
independently from $N(0, h/s_{\delta})$, while for each $j \notin S$, set $\delta_j = 0$.
\item Generate $\bm{\gamma}$. Let $\rho$ represent the proportion of influential points. Let $O$ be an index set with $|O| = \rho n$ randomly sampled from $\{1, . . . , n\}$ without replacement in target dataset and $O_{(k)}$ be an index set with $|O_{(k)}| = \rho N_{(k)}$ randomly sampled from $\{1, . . . , n_{(k)}\}$ without replacement in source dataset for $k=1,...,K$. Then for each $i\in O$, $\gamma_{i}$ follows a normal distribution $\mathcal{N}(a,b) $, where $a\sim U(0,20)$, $b \sim U(0,5)$. While for each $i \notin O$, set $\gamma_i=0$ otherwise. The setting of $\bm{\gamma}$ in the source datasets follows the same logic.
\end{itemize}
\end{Example}
\begin{figure}[ht]
\centering
\subfigure[$N=1000$]
    {\includegraphics[width=1.6in]{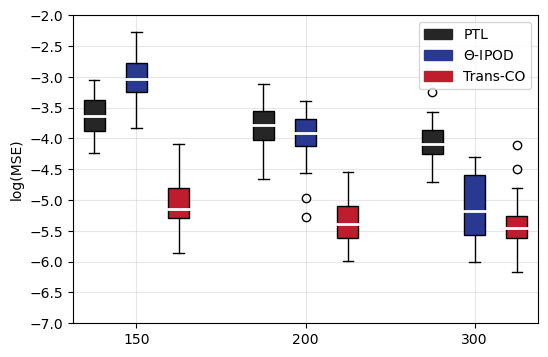}
    \label{fig:s15N1000o50}}
\hfil
\subfigure[$N=1500$]
    {\includegraphics[width=1.6in]{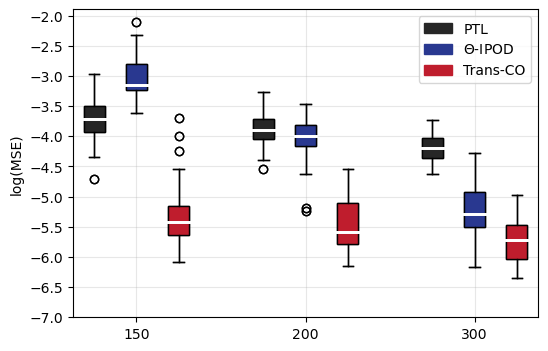}
    \label{fig:s15N1500o50}}
\hfil
\subfigure[$N=2000$]
    {\includegraphics[width=1.6in]{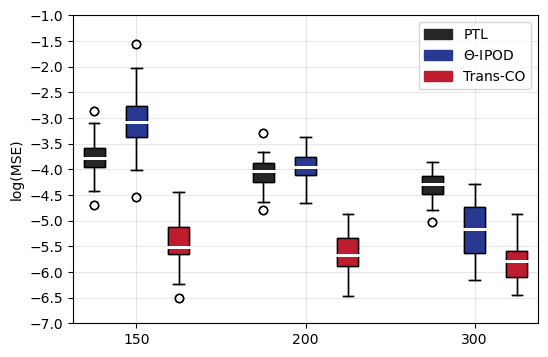}
    \label{fig:s15N2000o50}}
\caption{Comparison of different methods for different sample size of target dataset and source datasets when
$s = 25$ in Example \ref{ex1}.}
\label{figure:1}
\end{figure}

\begin{figure}[ht]
\centering
\subfigure[$N=1000$]
    {\includegraphics[width=1.6in]{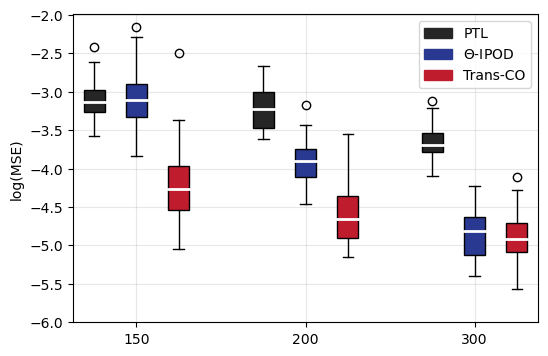}
    \label{fig:s75N1000o50}}
\hfil
\subfigure[$N=1500$]
    {\includegraphics[width=1.6in]{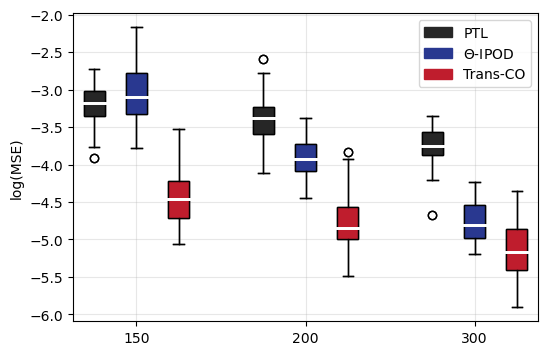}
    \label{fig:s75N1500o50}}
\hfil
\subfigure[$N=2000$]
    {\includegraphics[width=1.6in]{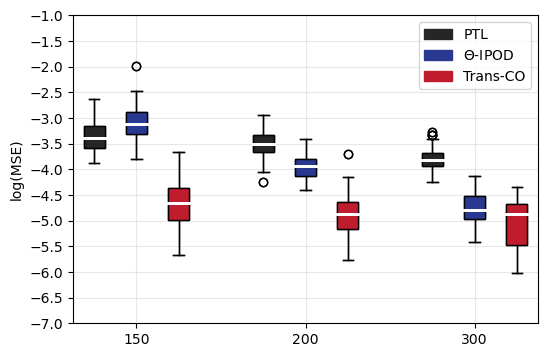}
    \label{fig:s75N2000o50}}
\caption{Comparison of different methods for different sample size of target dataset and source datasets when
$s = 75$ in Example \ref{ex1}.}
\label{figure:2}
\end{figure}

We conduct $50$ repeated experiments, and compute the Mean Squared Error (MSE) for the parameter $\bm{\beta}$ and the F1-score of influential point detection in each trial. To compare the performance of different methods and investigate their relationship with sample size, we plotted boxplots with logarithmic mean squared error shown as log(MSE) on the $y$-axis and the sample size of the target dataset on the $x$-axis. Additionally, we varied the sample size of the source dataset. The boxplots in Fig. \ref{figure:1} and Fig. \ref{figure:2} illustrate how the parameter estimation performance of each method changes with sample size. As the target sample size increases, the log(MSE) values decrease for all methods. However, our method consistently achieves the lowest log(MSE) values, indicating that our method yields parameter estimates closest to the true values.

\begin{table}[ht]   
\caption{Evaluation results of different algorithms. The results include the mean and standard error of F1-score ($\%$) for influential point detection on the target dataset in Example \ref{ex1}.}
\scriptsize
\label{table:1} 
\begin{tabular}{c c c c c c} 
\toprule
  \multirow{2}{*}{$N$}&\multirow{2}{*}{$n$}& \multicolumn{2}{c}{$s=25$} & \multicolumn{2}{c}{$s=75$}  \\ 
& & $\Theta$-IPOD &Trans-CO& $\Theta$-IPOD &Trans-CO\\
\midrule
\multirow{3}{*}{$1000$}&$150$&$34.37\pm8.08$&$79.85\pm15.55$&$35.60\pm8.81$&$67.82\pm24.34$\\
&$200$&$38.54\pm14.91$&$77.59\pm14.98$&$39.80\pm13.68$&$79.88\pm14.01$\\
&$300$&$58.34\pm25.08$&$83.59\pm8.94$&$59.92\pm23.02$&$80.19\pm9.11$\\
\midrule
\multirow{3}{*}{$1500$}&$150$&$33.78\pm6.68$&$82.11\pm14.74$&$35.55\pm7.61$&$69.40\pm22.26$\\
&$200$&$37.67\pm13.69$&$79.66\pm12.00$&$38.69\pm14.02$&$82.08\pm9.28$\\
&$300$&$63.99\pm22.53$&$85.33\pm7.79$&$53.62\pm24.46$&$81.12\pm11.77$\\
\midrule
\multirow{3}{*}{$2000$}&$150$&$36.54\pm10.23$&$83.02\pm16.34$&$35.92\pm7.52$&$77.81\pm16.01$\\
&$200$&$37.83\pm10.98$&$82.13\pm11.45$&$40.58\pm13.97$&$80.03\pm14.40$\\
&$300$&$57.51\pm22.64$&$84.64\pm7.99$&$55.68\pm22.97$&$80.05\pm10.65$\\
\bottomrule
\end{tabular} 
\end{table}

In addition, Table \ref{table:1} shows the results of anomaly detection accuracy, which include the average F1-score and the standard deviation of F1-score. Due to PTL does not have the function of detecting impact points, we only compare the performance of $\Theta$-IPOD and Trans-CO methods. Since we set $\rho=0.05$, we can find the index of $\bm{\gamma}$ that is not equal to $0$, which determine the true influential points and the detected influential points, and calculate F1-score accordingly. Through in-depth analysis of the experimental results in Table \ref{table:1}, it can be found that our proposed method exhibits significant advantages, specifically in having a higher average F1-score and a smaller F1-score standard deviation. The average F1-score, as the harmonic average of precision and recall, the higher its value, the stronger the comprehensive detection ability of this method in accurately identifying the influential points (precision) and finding as many true influential points as possible (recall). The standard deviation of F1-score reflects the degree of fluctuation in F1-score under different experiments or samples. The smaller the standard deviation, the more stable the detection performance of the method can be maintained in various situations. Combining these two aspects, it is sufficient to prove that our method Trans-CO has better performance in detecting influential points.

\begin{Example}
\label{ex2}
In this example, we focus on investigating the impact of the number of source datasets $K$ and influential point proportion $\rho$ on parameter estimation for different methods.
\begin{itemize}
    \item Set the number of source datasets $K\in \{2,4,6,8,10\}$, the weight vector $\bm{w}=(w_1,...,w_K)$, where $w_k$ is generated from uniform distribution $U(-2,2)$ for $k=1,..,K$. We consider the sample size of all source datasets $N_{(k)}=N=1000$ for $k=1,..,K$, and the sample size of target dataset $n=200$. Moreover, set the number of non-zero regression coefficients $s\in\{30,70\}$.
    \item Influential points will occur at a rate of $\rho\in\{0.01,0,05\}$ in both the source and target datasets, the configurations of the remaining parameters are identical to those in Example \ref{ex1}.
\end{itemize}
\end{Example}

\begin{figure}[ht]
\centering
\subfigure[$\rho=0.05$]
    {\includegraphics[width=2.5in]{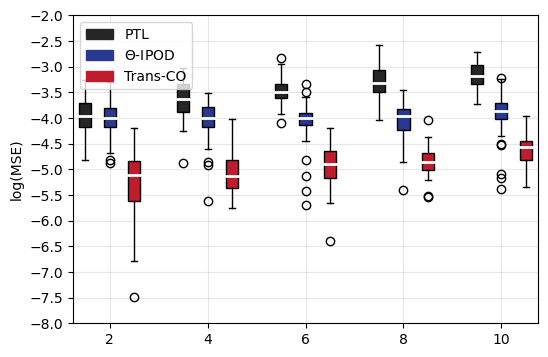}
    \label{fig:K_o30 10 0.01}}
\hfil
\subfigure[$\rho=0.1$]
    {\includegraphics[width=2.5in]{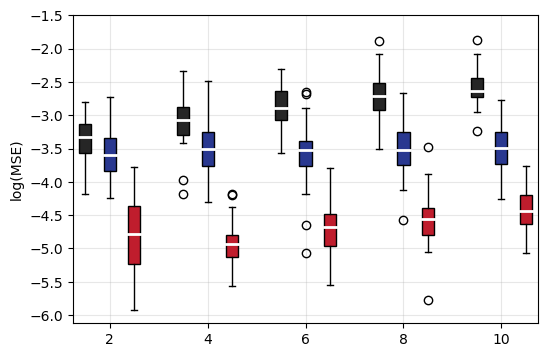}
    \label{fig:K_o30 10 0.05}}
\caption{Comparison of different methods for varies $K$ and $\rho$ when $s = 30$ in Example \ref{ex2}.}
\label{figure:3}
\end{figure}
\begin{figure}[ht]
\centering
\subfigure[$\rho=0.05$]
    {\includegraphics[width=2.5in]{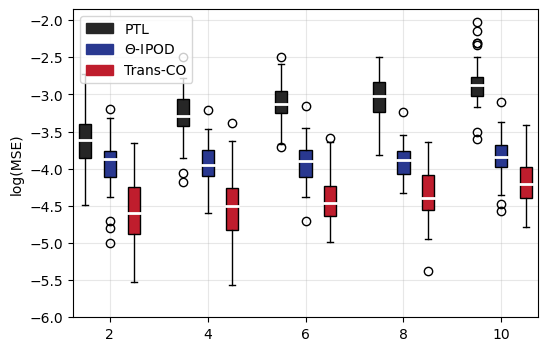}
    \label{fig:K_o70 10 0.01}}
\hfil
\subfigure[$\rho=0.1$]
    {\includegraphics[width=2.5in]{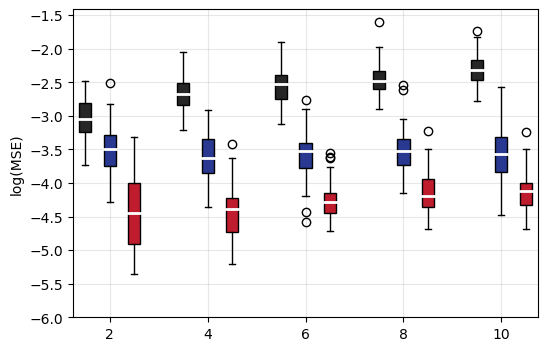}
    \label{fig:K_o70 10 0.05}}
\caption{Comparison of different methods for varies $K$ and $\rho$ when $s = 70$ in Example \ref{ex2}.}
\label{figure:4}
\end{figure}

A comparison of parameter estimation performance across various methods under different values of $K$ and $\rho$ are illustrated in Fig. \ref{figure:3} and \ref{figure:4}.
When the datasets contain a certain proportion of influential points, the log(MSE) of PTL and Trans-CO present a slight upward trend with the growth of the parameter $K$. Specifically, as the source model pool becomes more abundant, our method can maintain consistent and stable performance. Furthermore, as the influential point proportion grows, our method demonstrates superior performance in parameter estimation, achieving the minimal log(MSE) compared to alternative approaches. This indicates that, even with a substantial presence of influential points, our method effectively integrates valuable information from the source data and accurately transfers it to the target model. Consequently, our method exhibits greater applicability and advantages in transfer learning tasks under noisy conditions with abundant influential points.

\begin{table}[ht]   
\caption{Evaluation results of different algorithms. The results include the mean and standard error of F1-score ($\%$) for influential point detection on the target dataset in Example \ref{ex2}.}
\scriptsize
\label{table:2} 
\begin{tabular}{c c c c c c} 
\toprule
  \multirow{2}{*}{$\rho$}&\multirow{2}{*}{$K$}& \multicolumn{2}{c}{$s=30$} & \multicolumn{2}{c}{$s=70$}  \\ 
& & $\Theta$-IPOD &Trans-CO& $\Theta$-IPOD &Trans-CO\\
\midrule
\multirow{5}{*}{$0.05$}&$2$&$37.98\pm13.61$&$80.58\pm11.71$&$48.49\pm8.82$&$80.31\pm9.66$\\
&$4$&$39.65\pm15.21$&$79.55\pm11.24$&$49.53\pm11.22$&$83.20\pm7.47$\\
&$6$&$41.47\pm16.54$&$73.23\pm14.39$&$47.19\pm9.76$&$82.08\pm8.07$\\
&$8$&$40.31\pm16.46$&$80.90\pm11.97$&$47.59\pm7.41$&$82.11\pm10.09$\\
&$10$&$38.10\pm14.41$&$79.92\pm11.87$&$48.89\pm9.56$&$81.44\pm10.35$\\
\midrule
\multirow{5}{*}{$0.1$}&$2$&$48.49\pm8.82$&$80.31\pm9.66$&$46.58\pm7.80$&$81.02\pm8.95$\\
&$4$&$49.53\pm11.22$&$83.20\pm7.47$&$50.69\pm11.07$&$82.14\pm9.39$\\
&$6$&$47.19\pm9.76$&$82.08\pm8.07$&$49.06\pm8.05$&$78.27\pm9.49$\\
&$8$&$47.59\pm7.41$&$82.11\pm10.09$&$48.71\pm9.36$&$81.32\pm10.46$\\
&$10$&$48.89\pm9.56$&$81.44\pm10.35$&$47.68\pm6.81$&$79.13\pm10.24$\\
\bottomrule
\end{tabular} 
\end{table}

In addition to the preceding analysis, we also conducted a comprehensive study on the impact of varying numbers of source models and different proportions of influential points on the model's ability to detect influential points in Table \ref{table:2}. As can be clearly observed from the table, our proposed method consistently outperforms the $\Theta$-IPOD method under all circumstances. Moreover, with the increase in the proportion of influential points, our method exhibits a rising trend in detection accuracy, demonstrating its robustness and effectiveness in handling different scenarios.

\begin{Example}
\label{ex3}
In this example, the covariance matrices of covariates for the target dataset and the source datasets are different, and their error variances also differ. The generation and setting of the remaining parameters is the same as in Example \ref{ex1}.
\end{Example}
\begin{itemize}
    \item Generate $\Sigma$ and $\Sigma_{(k)}$. For the target dataset, the covariance matrix of the covariates is set to the identity matrix $\Sigma=\bm{I}_p$. For each source dataset, the covariance matrix $\Sigma_{(k)}$ is defined as a symmetric Toeplitz matrix, with its first row structured as $(1, \frac{\bm{1}_{2k-1}}{k+1},\bm{0}_{p-2k})$.
    \item Generate $\sigma^2$ and $\sigma_{(k)}^2$. Set the error variance for target dataset as $\sigma^2=1$, and the error variance for each source data as $\sigma_{(k)}^2 = \frac{k+1}{10}$.
\end{itemize}

\begin{figure}[ht]
\centering
\subfigure[$N=1000$]
    {\includegraphics[width=1.6in]{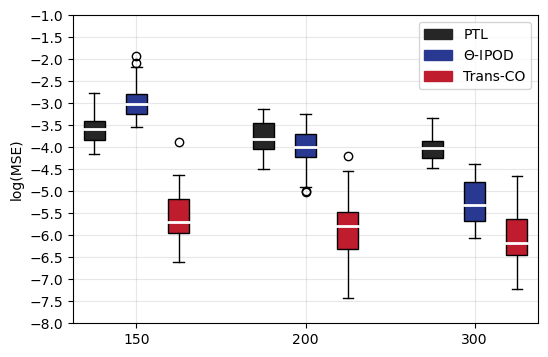}
    \label{fig:ex3s25N1000o50}}
\hfil
\subfigure[$N=1500$]
    {\includegraphics[width=1.6in]{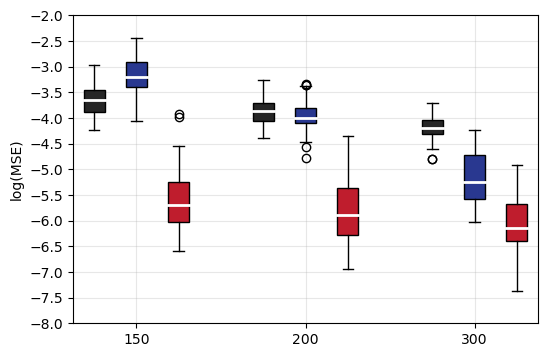}
    \label{fig:ex3s25N1500o50}}
\hfil
\subfigure[$N=2000$]
    {\includegraphics[width=1.6in]{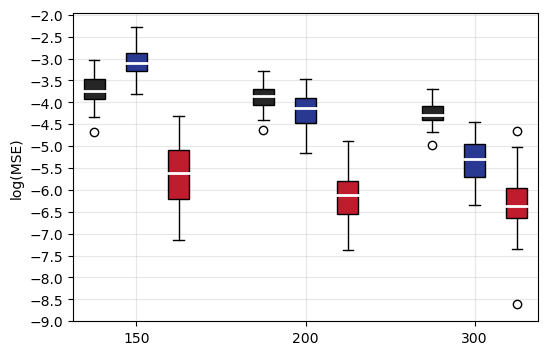}
    \label{fig:ex3s25N2000o50}}
\caption{Comparison of different methods for different sample size of target dataset and source datasets when
$s = 25$ in Example \ref{ex3}.}
\label{figure:5}
\end{figure}

\begin{figure}[ht]
\centering
\subfigure[$N=1000$]
    {\includegraphics[width=1.6in]{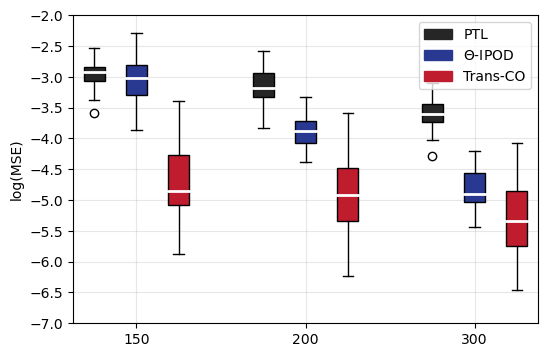}
    \label{fig:ex3s75N1000o50}}
\hfil
\subfigure[$N=1500$]
    {\includegraphics[width=1.6in]{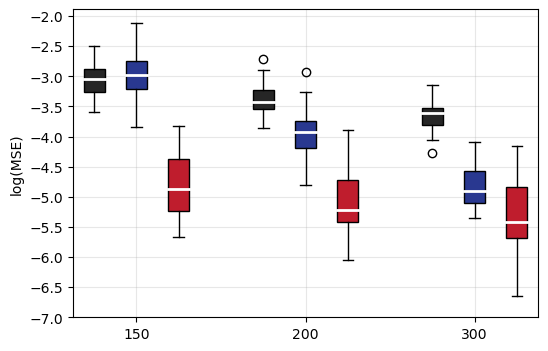}
    \label{fig:ex3s75N1500o50}}
\hfil
\subfigure[$N=2000$]
    {\includegraphics[width=1.6in]{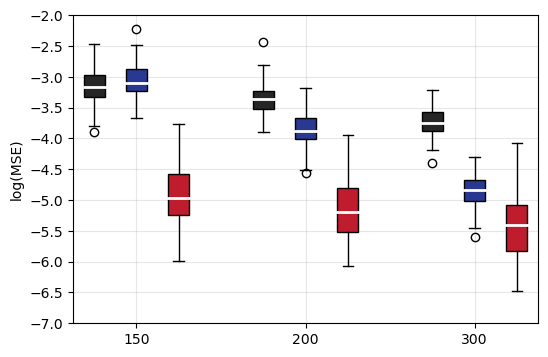}
    \label{fig:ex3s75N2000o50}}
\caption{Comparison of different methods for different sample size of target dataset and source datasets when
$s = 75$ in Example \ref{ex3}.}
\label{figure:6}
\end{figure}

\begin{table}[ht]   
\caption{Evaluation results of different algorithms. The results include the mean and standard error of F1-score ($\%$) for influential point detection on the target dataset in Example \ref{ex3}.}
\scriptsize
\label{table:3} 
\begin{tabular}{c c c c c c} 
\toprule
  \multirow{2}{*}{$N$}&\multirow{2}{*}{$n$}& \multicolumn{2}{c}{$s=25$} & \multicolumn{2}{c}{$s=75$}  \\ 
& & $\Theta$-IPOD &Trans-CO& $\Theta$-IPOD &Trans-CO\\
\midrule
\multirow{3}{*}{$1000$}&$150$&$34.36\pm6.69$&$79.73\pm14.17$&$36.01\pm7.87$&$79.37\pm16.28$\\
&$200$&$39.30\pm15.31$&$80.45\pm12.43$&$35.28\pm10.26$&$79.97\pm11.88$\\
&$300$&$59.04\pm23.20$&$85.05\pm9.37$&$59.19\pm23.29$&$81.55\pm8.85$\\
\midrule
\multirow{3}{*}{$1500$}&$150$&$36.95\pm9.71$&$83.51\pm12.05$&$35.38\pm8.60$&$75.05\pm18.06$\\
&$200$&$34.20\pm10.94$&$82.18\pm10.40$&$38.67\pm16.41$&$77.23\pm14.32$\\
&$300$&$63.51\pm22.55$&$85.17\pm8.26$&$57.42\pm24.13$&$82.29\pm12.34$\\
\midrule
\multirow{3}{*}{$2000$}&$150$&$36.73\pm7.72$&$82.00\pm13.09$&$35.57\pm8.21$&$80.06\pm13.01$\\
&$200$&$42.24\pm15.02$&$83.10\pm8.68$&$39.12\pm15.81$&$79.37\pm15.24$\\
&$300$&$57.87\pm22.89$&$86.10\pm6.53$&$57.87\pm22.89$&$83.34\pm8.66$\\
\bottomrule
\end{tabular} 
\end{table}

The comparison results of various methods under heteroscedasticity are shown in Figs. \ref{figure:5}, \ref{figure:6} and Table \ref{table:3}. Our Trans-CO method performs the best in parameter estimation when the feature data in each dataset is heteroscedastic and the error is also heteroscedastic.

\begin{Example}
\label{ex4}
    In this example, we assume that conditions of the unique identification parameters in linear approximation assumption do not hold, specifically, the condition $\bm{B}^{\top}\bm{\Sigma}\bm{\delta}=0$ is not satisfied.
\end{Example}
\begin{itemize}
    \item Generate $\delta$ and $\Sigma$. S randomly samples from $\{1,...,p\}$ without replacement. Set $\Sigma = (\sigma_{ij})_{p\times p}$ where $\sigma_{ij} = 0.5^{|i-j|}$. The generation of these two parameters cause the unique identification condition to not be met, and all other generation steps and parameter settings are the same as Example \ref{ex1}.
\end{itemize}

\begin{figure}[ht]
\centering
\subfigure[$N=1000$]
    {\includegraphics[width=1.6in]{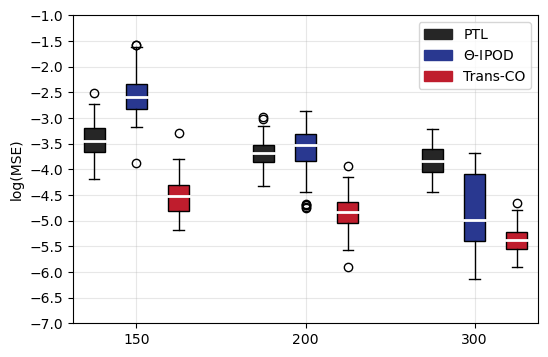}
    \label{fig:ex4s25N1000o50}}
\hfil
\subfigure[$N=1500$]
    {\includegraphics[width=1.6in]{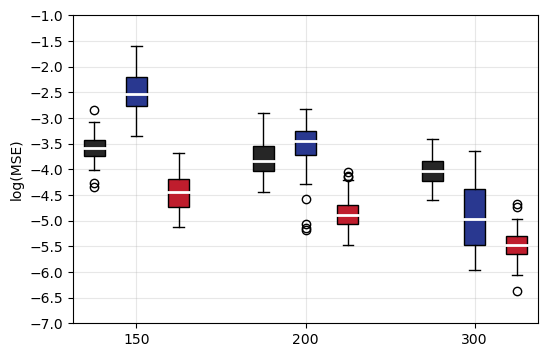}
    \label{fig:ex4s25N1500o50}}
\hfil
\subfigure[$N=2000$]
    {\includegraphics[width=1.6in]{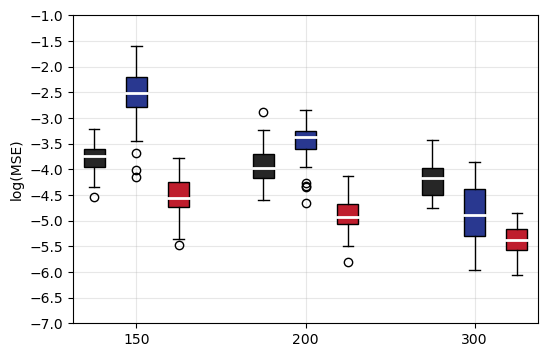}
    \label{fig:ex4s25N2000o50}}
\caption{Comparison of different methods for different sample size of target dataset and source datasets when
$s = 25$ in Example \ref{ex4}.}
\label{figure:7}
\end{figure}

\begin{figure}[ht]
\centering
\subfigure[$N=1000$]
    {\includegraphics[width=1.6in]{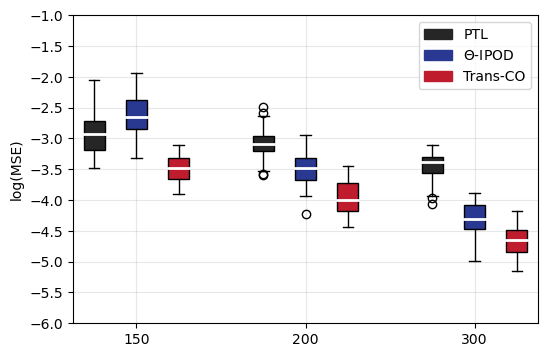}
    \label{fig:ex4s75N1000o50}}
\hfil
\subfigure[$N=1500$]
    {\includegraphics[width=1.6in]{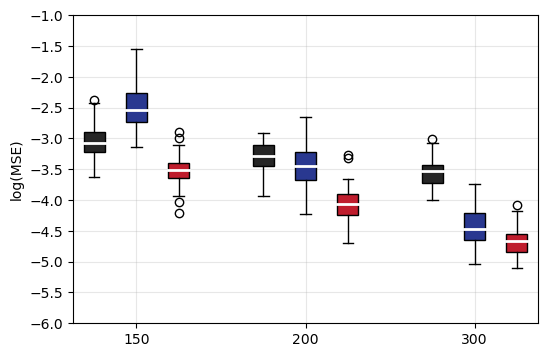}
    \label{fig:ex4s75N1500o50}}
\hfil
\subfigure[$N=2000$]
    {\includegraphics[width=1.6in]{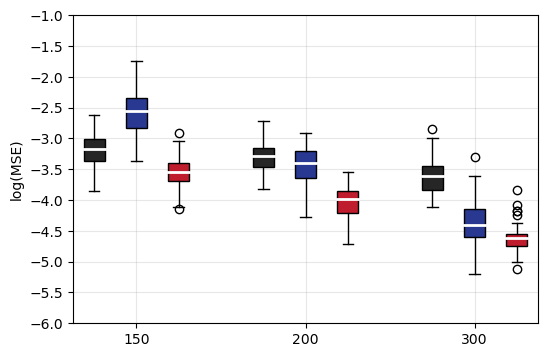}
    \label{fig:ex4s75N2000o50}}
\caption{Comparison of different methods for different sample size of target dataset and source datasets when
$s = 75$ in Example \ref{ex4}.}
\label{figure:8}
\end{figure}

\begin{table}[ht]   
\caption{Evaluation results of different algorithms. The results include the mean and standard error of F1-score ($\%$) for influential point detection on the target dataset in Example \ref{ex4}.}
\scriptsize
\label{table:4} 
\begin{tabular}{c c c c c c} 
\toprule
  \multirow{2}{*}{$N$}&\multirow{2}{*}{$n$}& \multicolumn{2}{c}{$s=25$} & \multicolumn{2}{c}{$s=75$}  \\ 
& & $\Theta$-IPOD &Trans-CO& $\Theta$-IPOD &Trans-CO\\
\midrule
\multirow{3}{*}{$1000$}&$150$&$37.15\pm10.25$& $82.28\pm15.39$ &$34.49\pm8.70$&$79.61\pm14.26$\\
&$200$&$39.57\pm13.54$&$81.85\pm10.89$&$40.19\pm17.11$&$85.23\pm7.90$\\
&$300$&$63.02\pm23.27$&$88.14\pm5.75$&$51.74\pm20.63$&$84.66\pm9.14$\\
\midrule
\multirow{3}{*}{$1500$}&$150$&$34.17\pm7.07$& $82.23\pm14.32$ &$33.55\pm7.55$&$78.92\pm10.29$\\
&$200$&$38.83\pm14.36$&$84.27\pm10.28$&$41.59\pm16.79$&$85.90\pm8.36$\\
&$300$&$64.29\pm20.93$&$86.30\pm10.67$&$60.71\pm22.76$&$85.83\pm5.99$\\
\midrule
\multirow{3}{*}{$2000$}&$150$&$36.56\pm11.60$& $82.31\pm16.77$ &$36.88\pm9.83$&$82.45\pm11.04$\\
&$200$&$36.51\pm12.73$&$84.03\pm8.01$&$38.71\pm13.29$&$84.24\pm10.44$\\
&$300$&$62.04\pm23.18$&$84.48\pm9.75$&$60.68\pm22.93$&$86.22\pm6.70$\\
\bottomrule
\end{tabular} 
\end{table}
Despite the fact that the unique identification conditions in the linear approximation assumption do not hold, our Trans-CO method still performs the small log(MSE) and the narrow interquartile ranges (IQRs), as demonstrated in Figs. \ref{figure:7} and \ref{figure:8}. In addition, it shows superior performance in the detection of influential points in Table \ref{table:4}.

\begin{Example}
\label{ex5}
    In this example, we systematically compare methods in high-dimensional settings where $n<p$.
\end{Example}
\begin{itemize}
    \item We set $n\in\{10,30,50,70,90\}$, and all other generation steps and parameter settings are the same as Example \ref{ex1}.
\end{itemize}

\begin{figure}[ht]
\centering
\subfigure[$N=1000$]
    {\includegraphics[width=1.6in]{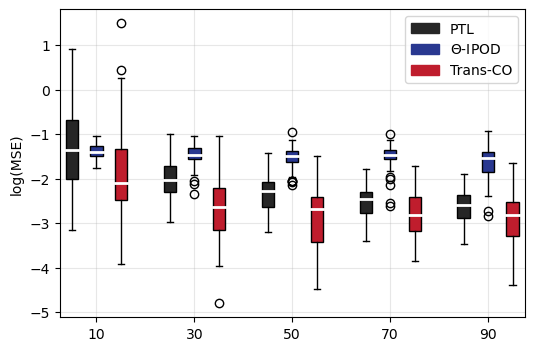}
    \label{fig:ex5s25N1000o50}}
\hfil
\subfigure[$N=1500$]
    {\includegraphics[width=1.6in]{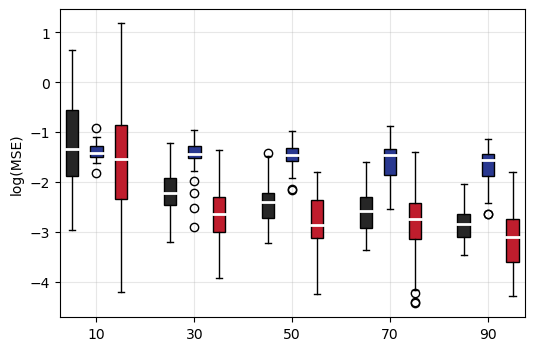}
    \label{fig:ex5s25N1500o50}}
\hfil
\subfigure[$N=2000$]
    {\includegraphics[width=1.6in]{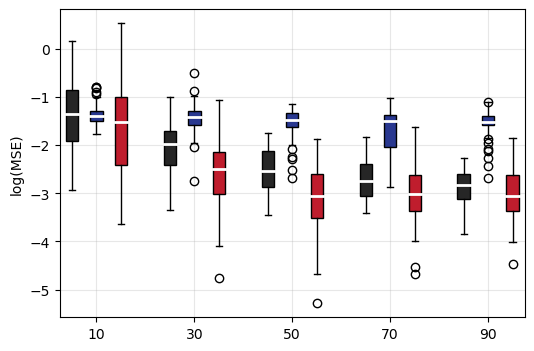}
    \label{fig:ex5s25N2000o50}}
\caption{Comparison of different methods for different sample size of target dataset and source datasets when
$s = 25$ in Example \ref{ex5}.}
\label{figure:9}
\end{figure}
\begin{figure}[ht]
\centering
\subfigure[$N=1000$]
    {\includegraphics[width=1.6in]{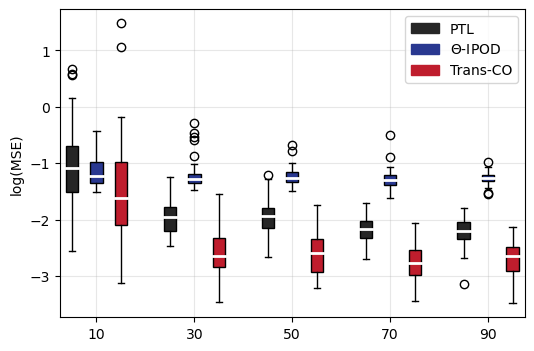}
    \label{fig:ex5s75N1000o50}}
\hfil
\subfigure[$N=1500$]
    {\includegraphics[width=1.6in]{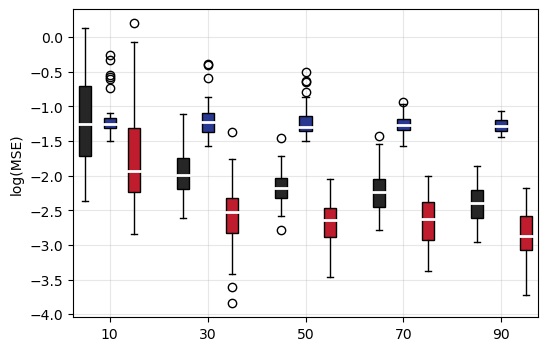}
    \label{fig:ex5s75N1500o50}}
\hfil
\subfigure[$N=2000$]
    {\includegraphics[width=1.6in]{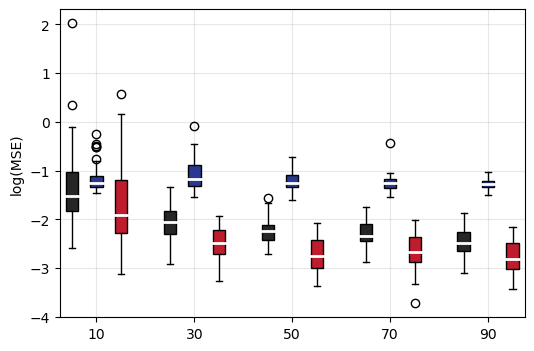}
    \label{fig:ex5s75N2000o50}}
\caption{Comparison of different methods for different sample size of target dataset and source datasets when
$s = 75$ in Example \ref{ex5}.}
\label{figure:10}
\end{figure}

\begin{table}[ht]   
\caption{Evaluation results of different algorithms. The results include the mean and standard error of F1-score ($\%$) for influential point detection on the target dataset in Example \ref{ex5}.}
\scriptsize
\label{table:5} 
\begin{tabular}{c c c c c c} 
\toprule
  \multirow{2}{*}{$N$}&\multirow{2}{*}{$n$}& \multicolumn{2}{c}{$s=25$} & \multicolumn{2}{c}{$s=75$}  \\ 
& & $\Theta$-IPOD &Trans-CO& $\Theta$-IPOD &Trans-CO\\
\midrule
\multirow{5}{*}{$1000$}&$10$&$37.33\pm28.78$& $37.00\pm43.37$ &$25.67\pm27.73$&$34.67\pm42.93$\\
&$30$&$50.98\pm19.74$&$74.05\pm22.37$&$34.31\pm22.43$&$60.90\pm28.62$\\
&$50$&$47.03\pm19.24$&$74.25\pm15.76$&$39.01\pm19.24$&$64.15\pm17.12$\\
&$70$&$52.64\pm16.39$&$70.75\pm15.92$&$49.97\pm19.80$&$75.66\pm15.72$\\
&$90$&$53.92\pm17.73$&$77.89\pm12.46$&$53.41\pm18.17$&$76.96\pm12.57$\\
\midrule
\multirow{5}{*}{$1500$}&$10$&$36.67\pm28.87$& $36.67\pm42.03$ &$30.00\pm29.44$&$44.67\pm47.78$\\
&$30$&$40.00\pm24.13$&$64.15\pm24.69$&$38.20\pm20.74$&$74.90\pm18.31$\\
&$50$&$49.64\pm18.70$&$75.85\pm17.12$&$39.69\pm20.47$&$74.92\pm16.59$\\
&$70$&$42.91\pm19.14$&$75.95\pm14.44$&$47.59\pm20.42$&$78.94\pm11.70$\\
&$90$&$46.39\pm21.72$&$77.26\pm11.40$&$47.59\pm20.42$&$78.94\pm11.70$\\
\midrule
\multirow{5}{*}{$2000$}&$10$&$38.00\pm28.68$& $46.67\pm44.35$ &$31.00\pm30.37$&$46.00\pm44.91$\\
&$30$&$38.66\pm25.95$&$70.22\pm26.66$&$30.39\pm22.28$&$63.53\pm26.15$\\
&$50$&$47.91\pm19.06$&$74.88\pm14.84$&$46.94\pm22.28$&$74.47\pm16.21$\\
&$70$&$45.37\pm19.43$&$78.41\pm13.86$&$40.40\pm23.59$&$77.22\pm12.54$\\
&$90$&$56.57\pm16.68$&$77.47\pm13.92$&$50.69\pm19.14$&$76.97\pm13.28$\\
\bottomrule
\end{tabular} 
\end{table}

In Example \ref{ex5}, we investigate high-dimensional settings where the sample size of target dataset ($n$) is smaller than the feature dimension ($p$). The IPOD algorithm employs extended method in She (\citeyear{she2011outlier}) for high-dimensional adaptation, while PTL and Trans-CO are inherently suitable for such scenarios. Our proposed method demonstrates superior performance in two key aspects as demonstrated in Figs. \ref{figure:9}, \ref{figure:10} and Table \ref{table:5}. The mean of log(MSE) of the estimated regression coefficients ($\hat{\bm{\beta}}$) achieved by our method is the lowest among all compared algorithms. Compared to IPOD, our method attains a higher average F1-score with smaller standard deviation, reflecting both improved accuracy and stability in influential point detection. Furthermore, the detection accuracy improves consistently as $n$ increases.

\section{Experiments on real data}\label{Experiments on Real Data}
In this study, we employ Beijing Multi-Site Air Quality \footnote{\href{https://archive.ics.uci.edu/dataset/501/beijing+multi+site+air+quality+data}{https://archive.ics.uci.edu/dataset/501/beijing+multi+site+air+quality+data}} as a real dataset to further evaluate the proposed methodology. The Air Quality dataset, covering March $2013$ to February $2017$, is from the Beijing Municipal Environmental Monitoring Center. In addition to the temporal information and Nominal Variable, the dataset comprises 11 variables, including PM2.5, PM10, SO2, NO2, CO, O3, TEMP, PRES, DEWP, RAIN, WSPM. Given that the variable RAIN has a $0$ value rate exceeding $95\%$, it is excluded from further analysis to ensure model quality. This study selects co as the response variable, aiming to explore the potential influence of the remaining 9 feature variables. During the data preprocessing stage, all samples containing missing values are removed. The remaining 10 variables are then standardized to eliminate dimensional inconsistencies. 


The dataset including 320022 samples is divided into subsets based on the station in Beijing. Considering that transfer learning is often applied in scenarios where the target dataset is relatively small, we randomly select a portion of samples from each dataset for experimentation. Specifically, $5\%$ of the samples are randomly selected from each dataset of Aotizhongxin, Changping, Dingling, Dongsi, Guanyuan, Gucheng, Huairou, Nongzhanguan and Shunyi stations. These selected samples serve as 9 source domains to facilitate knowledge transfer. While only $1$\text{\textperthousand} of the samples from Tiantan station are selected as the target domain for evaluating transfer learning performance. During the training and parameter tuning phase, the target domain data is further split into $70\%$ for training and $30\%$ for testing.
Given that this study focuses on assessing the performance of transfer learning in the presence of influential points, we use the same proportion as identifying influential points in the training set to remove influential points from the test set. A total of 500 experiments are conducted to ensure reliability of the results.

\begin{figure}[ht]
\centering
\subfigure[Log(L$_\alpha$)]
    {\includegraphics[width=2in]{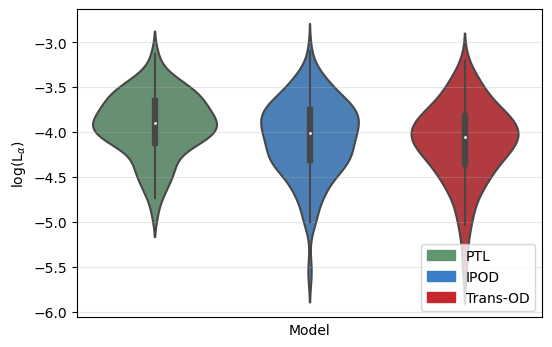}
    \label{fig:huber1}}
\hfil
\subfigure[R-squared]
    {\includegraphics[width=2in]{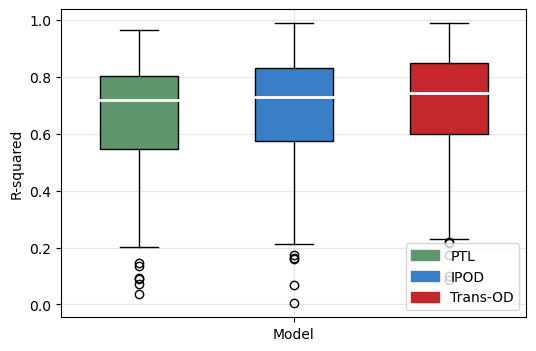}
    \label{fig:huber2}}
\caption{Comparison of the predicted Log(L$_\alpha$) and R-squared on the test set of the target dataset.}
\label{figure:huber}
\end{figure}

For model evaluation, we employed the three models and assessed their fitting performance and predictive performance separately on the test set of the target dataset. We use the Huber Loss (L$_\alpha$) and the coefficient of determination (R-squared) as evaluation metrics. The formula for L$_\alpha$ is as follows:
\begin{equation}
    L_\alpha = \frac{1}{n}\sum\limits_{i=1}^{n}l_\alpha(y_i,\hat{y_i}),
\end{equation}
where $l_\alpha(y_i-\hat{y_i}) = 
\begin{cases}
\frac{1}{2}(y_i-\hat{y_i})^2, & \mbox{$|y_i-\hat{y_i}|\leq\alpha$}, \\
\alpha|y_i-\hat{y_i}|-\frac{1}{2}\alpha^2, & \mbox{$|y_i-\hat{y_i}|>\alpha$}.
\end{cases}$ $\alpha$ is a threshold that determines when the loss function switches from quadratic to linear loss, and we set $\alpha=0.05$. Huber Loss serves as a robust evaluation metric that is less sensitive to influential points, thereby mitigating the impact of potential influential points in the test set on the model's performance. And the formula for R-squared is calculated as:
\begin{equation}
    \text{R-squared}=1-\frac{\text{SSR}}{\text{SST}},
\end{equation}
where $\text{SSR} = \sum\limits_{i=1}^{n}(y_i-\hat{y_i})^2$, $\text{SST} = \sum\limits_{i=1}^{n}(y_i-\bar{y})^2$, and $\bar{y}$ is the mean of the actual observed values in target test set. This metric evaluates the model's fitting performance by comparing the variance explained by the model with the total variance inherent in the data. A value closer to $1$ indicates a better fit of the model to the data. We exclude individual cases where the R-squared is negative and subsequently plotted the experimental results in Fig. \ref{figure:huber}.

Fig. \ref{fig:huber1} illustrates that, the Log(L$_\alpha$) violin plot of our proposed Trans-CO method exhibits the smallest average error, outperforming both the PTL and $\Theta$-IPOD methods. The proportion of variance explained by the model relative to the total variance is illustrated by Fig. \ref{fig:huber2}. Although the average R-squared values for all three methods exceed $0.7$, our Trans-CO method achieves a predicted R-squared that is $9.5\%$ higher than that of the PTL and $2.4\%$ higher than that of the IPOD.


\section{Conclusion and discussion}\label{section:Conclusion and Discussion}
\subsection{Conclusion}
In domains such as finance, industry, and healthcare, data scarcity often stems from the high costs associated with data acquisition, and the collected data may also contain influential points. Addressing the degradation of fitting performance in high-dimensional regression models caused by insufficient data volume and the presence of influential points, this study innovatively incorporates transfer learning into a thresholding-based iterative procedure for influential point detection. By leveraging knowledge learned from a source domain, our approach enhances learning performance in the target domain while reducing reliance on labeled data for the target task. The proposed Trans-CO algorithm optimizes model fitting and simultaneously detects influential points under data-limited conditions. Furthermore, we theoretically prove the convergence of the objective function, which is empirically validated through extensive simulations. These simulations comprehensively explore the impact of varying parameters, including sample size, variable sparsity, drift ratio, and the number of source models, on transfer learning performance. Simulations are also conducted under scenarios of heteroscedasticity and violations of unique identification conditions. The Trans-CO method demonstrates superior performance in all cases, and show remarkable versatility in addressing both classical ($n>p$) and high-dimensional ($n<p$) regression scenarios. Additionally, a real-world case study on Beijing Multi-Site Air Quality prediction further confirms the outstanding predictive efficacy of our approach. In summary, traditional statistical learning methods often struggle with performance degradation due to data scarcity and the interference of influential points. In contrast, our proposed Trans-CO transfer learning algorithm, leveraging the flexibility of cross-domain knowledge reuse, provides researchers and practitioners with an efficient solution to tackle the challenges of data scarcity and influential point detection.
\subsection{Discussion}
Our method takes into account both the training process of the source model and the identification of influential points, thus increasing the algorithm's complexity. Although it is more computationally intensive compared to the other two ablation models, the proposed algorithm demonstrates extremely outstanding performance in numerical simulations and empirical studies. The additional time cost it incurs is marginal when compared to the benefits derived from the significant improvement in accuracy.
Besides, since the data volume in scenarios where transfer learning is applicable is inherently not large, the running time is acceptable and not a significant issue. If there are particularly high demands on computational costs, we also attempt to optimize the algorithm's runtime by leveraging interpolation techniques to adjust initial values during the iterative process, thereby accelerating convergence and reducing iteration counts. In addition, Yu et al. (\citeyear{Yu02012025}) extended subsampling techniques to the Akaike information criterion (AIC) and the smoothed AIC model-averaging framework for generalized linear models. Inspired by this, we can utilize similar subsampling techniques during the training phase in subsequent work to address computational challenges in massive datasets. This approach is feasible because, under the linear approximation assumption, the weighted process of parameter transfer is essentially a form of model averaging. In the future, we will also try to adopt the deterministic approach that minimizes the Kullback-Leibler divergence (Wang and Sun \citeyear{wang2024deterministic}) and adaptive subsampling with the minimum energy criterion (Dai et al. \citeyear{dai2023subsampling}) to extract representative points, thereby reducing the training time of the source model. Additionally, we aim to explore the applicability of our framework and theoretical analyses to other domains.

\backmatter

\bmhead{Data Availability} 
The Beijing Multi-Site Air Quality dataset used in this study is publicly available and can be accessed as follows: 

\href{https://archive.ics.uci.edu/dataset/501/beijing+multi+site+air+quality+data}{https://archive.ics.uci.edu/dataset/501/beijing+multi+site+air+quality+data}. The dataset is publicly accessible and can be utilized for further research under their respective terms of use.

\bmhead{Acknowledgements} 

This work is supported by the National Natural Science Foundation of China under Grant 12471246.

\bmhead{Declarations}
The authors declare that they have no conflict of interest.


\begin{appendices}
\section{Proof of Theorem \ref{thm:1}}\label{Proof1}
\begin{proof}
This proof is mainly followed by the proof of Theorem 4.1 of She and Owen (\citeyear{she2011outlier}). We need to prove the three inequalities in (\ref{equation objective function}). 

(1) The proof of the first inequality is as follows: Given $\bm{w}^{(i)}$, minimizing $f$ over $\bm{\xi}$:
\begin{equation}
     \bm{\xi}^{(i+1)}=\underset{\bm{\xi}}{argmin}\frac{1}{2}\|\bm{Y}-\bm{X}\hat{\bm{B}}\bm{w}^{(i)}-\bm{M\xi}\|_2^2+P(\bm{\xi};\lambda),
\end{equation}
which is equivalent to minimizing the following equation:
\begin{equation}
    g(\xi) = a\frac{(t-\xi)^2}{2}+P(\xi;\lambda).
\end{equation}
where $P(\xi;\lambda) = P(0;\lambda)+P_{\Theta}(\xi;\lambda)+q(\xi;\lambda)$, $q(\cdot;\lambda)$ is nonnegative and $q(\Theta(\xi;\lambda);\lambda)=0$ for all $\xi$. The generalization of Proposition 3.2 in Antoniadis (\citeyear{antoniadis2007wavelet}) shows that the above minimization problem has a unique optimal solution $\Theta(t;\lambda)$ for every $t$ at which $\Theta(\cdot;\lambda)$ is continuous. Suppose $t>0$ and $\xi>\Theta(t;\lambda)$ without loss of generality. It suffices to consider $\xi\geq0$ since $g(\xi)\leq g(-\xi)$, where $g(\xi) = (t-\xi)^2/2+P(\xi;\lambda)$. Note that $\Theta^{-1}(u;\lambda) = sup\{t:\Theta(t;\lambda)\leq u\}$, $s(u,\lambda) = \Theta^{-1}(u;\lambda)-u$, and $P_{\Theta}(\xi;\lambda) = \int_{0}^{|\xi|}s(u;\lambda)du$ Then,
\begin{equation}
\begin{aligned}
    g(\xi)-g(\Theta(t;\lambda)) & = \int_{\Theta(t;\lambda)}^{\xi}g'(u)du\\ 
    &= \int_{\Theta(t;\lambda)}^{\xi}(u-t+P'(u;\lambda))du\\
    & = \int_{\Theta(t;\lambda)}^{\xi}(u-t+P'_{\Theta}(u;\lambda))du+q(\xi;\lambda)-q(\Theta(\xi;\lambda);\lambda)\\
    & = \int_{\Theta(t;\lambda)}^{\xi}(u-t+\Theta^{-1}(u;\lambda)-u)du+q(\xi;\lambda)\\
    & = \int_{\Theta(t;\lambda)}^{\xi}(\Theta^{-1}(u;\lambda)-t)du+q(\xi;\lambda)\\
\end{aligned}
\end{equation}
By definition $\Theta^{-1}(u;\lambda) = sup\{t:\Theta(t;\lambda)\leq u\}$, we know $\Theta^{-1}(u;\lambda) \geq t$, and then $g(\xi)\geq g(\Theta(t;\lambda))$. A comparable line of reasoning holds for the scenario where $\xi \leq \Theta(t;\lambda)$.

(2) The proof of the second inequality is as follows: Given $\bm{\xi}^{(i)}$, minimizing $f$ over $\bm{w}$ is equivalent to minimizing the following equation:
\begin{equation}
     \bm{w}^{(i+1)}=\underset{\bm{w}}{argmin} \frac{1}{2}\|\bm{Y}-\bm{\gamma}^{(i)}-\bm{XBw} -\bm{X\delta}^{(i)}\|_2^2
\end{equation}

The proof is now complete.
\end{proof}

\end{appendices}
\bibliography{sn-bibliography}

\begin{thebibliography}{27}
\providecommand{\natexlab}[1]{#1}
\providecommand{\url}[1]{{#1}}
\providecommand{\urlprefix}{URL }
\providecommand{\doi}[1]{\url{https://doi.org/#1}}
\providecommand{\eprint}[2][]{\url{#2}}
 \bibcommenthead

\bibitem[{Abdallah et~al.(2023)Abdallah, Joung, Lee, Mousoulis, Raghunathan, Shakouri, Sutherland, and Bagchi}]{abdallah2023anomaly}
Abdallah M, Joung BG, Lee WJ, et~al (2023) Anomaly detection and inter-sensor transfer learning on smart manufacturing datasets. Sensors 23(1):486. \doi{10.3390/s23010486}

\bibitem[{Aguinis et~al.(2013)Aguinis, Gottfredson, and Joo}]{aguinis2013best}
Aguinis H, Gottfredson RK, Joo H (2013) Best-practice recommendations for defining, identifying, and handling outliers. Organ Res Methods 16(2):270--301. \doi{10.1177/1094428112470848}

\bibitem[{Antoniadis(2007)}]{antoniadis2007wavelet}
Antoniadis A (2007) Wavelet methods in statistics: some recent developments and their applications. Stat Surv 1:16 -- 55. \doi{10.1214/07-SS014}

\bibitem[{Belsley et~al.(2005)Belsley, Kuh, and Welsch}]{belsley2005regression}
Belsley DA, Kuh E, Welsch RE (2005) Regression diagnostics: Identifying influential data and sources of collinearity. John Wiley \& Sons

\bibitem[{Bottmer et~al.(2022)Bottmer, Croux, and Wilms}]{bottmer2022sparse}
Bottmer L, Croux C, Wilms I (2022) Sparse regression for large data sets with outliers. Eur J Oper Res 297(2):782--794. \doi{10.1016/j.ejor.2021.05.049}

\bibitem[{Chen and Song(2025)}]{chen2025transfer}
Chen X, Song Y (2025) Transfer learning for semiparametric varying coefficient spatial autoregressive models. Stat Pap 66(2):1--22. \doi{10.1007/s00362-025-01662-5}

\bibitem[{Cousineau and Chartier(2010)}]{cousineau2010outliers}
Cousineau D, Chartier S (2010) Outliers detection and treatment: a review. Int J Psychol Res 3(1):58--67

\bibitem[{Dai et~al.(2023)Dai, Song, and Wang}]{dai2023subsampling}
Dai W, Song Y, Wang D (2023) A subsampling method for regression problems based on minimum energy criterion. Technometrics 65(2):192--205. \doi{10.1080/00401706.2022.2127915}

\bibitem[{Jin et~al.(2024)Jin, Yan, Aseltine, and Chen}]{jin2024transfer}
Jin J, Yan J, Aseltine RH, et~al (2024) Transfer learning with large-scale quantile regression. Technometrics 66(3):381--393. \doi{10.1080/00401706.2024.2315952}

\bibitem[{Klivans et~al.(2018)Klivans, Kothari, and Meka}]{klivans2018efficient}
Klivans A, Kothari PK, Meka R (2018) Efficient algorithms for outlier-robust regression. In: Conference On Learning Theory, PMLR, pp 1420--1430

\bibitem[{Li et~al.(2022)Li, Cai, and Li}]{li2022transfer}
Li S, Cai TT, Li H (2022) Transfer learning for high-dimensional linear regression: Prediction, estimation and minimax optimality. J R Stat Soc B 84(1):149--173. \doi{10.1080/01621459.2022.2071278}

\bibitem[{Li et~al.(2024)Li, Zhang, Cai, and Li}]{li2024estimation}
Li S, Zhang L, Cai TT, et~al (2024) Estimation and inference for high-dimensional generalized linear models with knowledge transfer. J Am Stat Assoc 119(546):1274--1285. \doi{10.1080/01621459.2023.2184373}

\bibitem[{Lin et~al.(2024)Lin, Zhao, Wang, and Wang}]{lin2024profiled}
Lin Z, Zhao J, Wang F, et~al (2024) Profiled transfer learning for high dimensional linear model. arXiv preprint {\href{https://arxiv.org/abs/arXiv:2406.00701}{{arXiv:2406.00701}}}

\bibitem[{Liu et~al.(2020)Liu, Shen, Li, and Caramanis}]{liu2020high}
Liu L, Shen Y, Li T, et~al (2020) High dimensional robust sparse regression. In: International Conference on Artificial Intelligence and Statistics, PMLR, pp 411--421

\bibitem[{Lockner et~al.(2022)Lockner, Hopmann, and Zhao}]{lockner2022transfer}
Lockner Y, Hopmann C, Zhao W (2022) Transfer learning with artificial neural networks between injection molding processes and different polymer materials. J Manuf Process 73:395--408. \doi{10.1016/j.jmapro.2021.11.014}

\bibitem[{Lou and Yang(2025)}]{lou2025joint}
Lou D, Yang Y (2025) Joint estimation of transfer learning on time series data. Stat Pap 66(1):1--19. \doi{10.1007/s00362-024-01629-y}

\bibitem[{Pan et~al.(2023)Pan, Bao, and Li}]{pan2023transfer}
Pan Q, Bao Y, Li H (2023) Transfer learning-based data anomaly detection for structural health monitoring. Struct Health Monit 22(5):3077--3091. \doi{10.1177/14759217221142174}

\bibitem[{Panjapornpon et~al.(2023)Panjapornpon, Bardeeniz, Hussain, and Chomchai}]{panjapornpon2023explainable}
Panjapornpon C, Bardeeniz S, Hussain MA, et~al (2023) Explainable deep transfer learning for energy efficiency prediction based on uncertainty detection and identification. Energy And Ai 12:100224. \doi{10.1016/j.egyai.2022.100224}

\bibitem[{She(2009)}]{she2009thresholding}
She Y (2009) Thresholding-based iterative selection procedures for model selection and shrinkage. Electron J Stat 3:384--415. \doi{10.1214/08-EJS348}

\bibitem[{She and Owen(2011)}]{she2011outlier}
She Y, Owen AB (2011) Outlier detection using nonconvex penalized regression. J Am Stat Assoc 106(494):626--639. \doi{10.1198/jasa.2011.tm10390}

\bibitem[{Tian and Feng(2023)}]{tian2023transfer}
Tian Y, Feng Y (2023) Transfer learning under high-dimensional generalized linear models. J Am Stat Assoc 118(544):2684--2697

\bibitem[{Tripuraneni et~al.(2021)Tripuraneni, Jin, and Jordan}]{tripuraneni2021provable}
Tripuraneni N, Jin C, Jordan M (2021) Provable meta-learning of linear representations. In: International conference on machine learning, PMLR, pp 10434--10443

\bibitem[{Wang and Sun(2024)}]{wang2024deterministic}
Wang S, Sun F (2024) Deterministic sampling based on kullback--leibler divergence and its applications. Statistical Papers 65(3):1411--1436. \doi{10.1007/s00362-023-01449-6}

\bibitem[{Yan et~al.(2024)Yan, Abdulkadir, Luley, Rosenthal, Schatte, Grewe, and Stadelmann}]{yan2024comprehensive}
Yan P, Abdulkadir A, Luley PP, et~al (2024) A comprehensive survey of deep transfer learning for anomaly detection in industrial time series: Methods, applications, and directions. IEEE Access 12:3768--3789. \doi{10.1109/ACCESS.2023.3349132}

\bibitem[{Yao et~al.(2022)Yao, Ge, Yu, and Xie}]{yao2022model}
Yao Y, Ge D, Yu J, et~al (2022) Model-based deep transfer learning method to fault detection and diagnosis in nuclear power plants. Front Energy Res 10:823395. \doi{10.3389/fenrg.2022.823395}

\bibitem[{Yu et~al.(2025)Yu, Wang, and Ai}]{Yu02012025}
Yu J, Wang H, Ai M (2025) A subsampling strategy for aic-based model averaging with generalized linear models. Technometrics 67(1):122--132. \doi{10.1080/00401706.2024.2407310}

\bibitem[{Zabin et~al.(2023)Zabin, Choi, and Uddin}]{zabin2023hybrid}
Zabin M, Choi HJ, Uddin J (2023) Hybrid deep transfer learning architecture for industrial fault diagnosis using hilbert transform and dcnn--lstm. J Supercomput 79(5):5181--5200. \doi{10.1007/s11227-022-04830-8}

\end{thebibliography}

\end{document}